\documentclass[12pt,preprint]{aastex}
\usepackage{graphicx}
\usepackage{dcolumn}
\usepackage{bm}

\begin{document}
\title{Cosmological Flashes from Rotating Black Holes}
\author{Maurice H.P.M. van Putten}
\affil{LIGO Laboratory, MIT 17-161, 175 Albany Street, Cambridge, MA 02139}

\begin{abstract}

  Current phenomenology suggests the presence of a compact baryon-poor energy
  source to cosmological gamma-ray bursts reacting to high-density matter.
  The association of short and long gamma-ray bursts with hyper- and suspended 
  accretion onto slowly and rapidly spinning black holes predicts weak X-ray 
  afterglow emissions from short bursts, as in GRB050509B and GRB050709. 
  Long gamma-ray bursts are probably the birth place of rapidly spinning high-mass 
  black holes in core-collapse of massive stars, as in GRB030329 with supernova
  SN2003dh. This predicts a long-duration burst of gravitational radiation powered by
  the spin-energy of the black hole. In contrast to MeV-neutrino emissions, as in SN1987A, 
  this can be tested by the advanced detectors LIGO and Virgo about once per year 
  up to distances of about 100Mpc. Detection of the expected 
  chirps and long-duration bursts of gravitational waves promises identification of
  Kerr black holes as the most luminous objects in the Universe. 
  
\end{abstract}
\maketitle

\section{Introduction}

Gammma-ray bursts (GRBs) were discovered serendipitously as ``flashes in the sky"
by the nuclear test-ban monitoring satellites Vela (US) and Konus (USSR).
These were soon recognized to be a natural phenomenon. The first data were 
publicly released by \cite{kle73} and \cite{maz74}.
The mysterious origin of this new astrophysical transient, lasting up to tens of seconds
in non-thermal gamma-ray radiation, gradually triggered 
a huge interest from the high-energy astrophysics community. The 
Burst and Transient Source Experiment (BATSE) on NASA's Compton Gamma-Ray
Observatory, launched in 1991, identified an isotropic distribution
in the sky and revealed the existence of short and long gamma-ray bursts, whose 
durations are broadly distributed around 0.3s and 30s, respectively 
\citep{kou93,pac99}. Moreover, the sensitivity of BATSE showed a deficit in
faint bursts in the number-versus-intensity distribution distinct from a -3/2
powerlaw. \cite{mee92} hereby established that gamma-ray bursts
are of cosmological origin. More recently, the Italian-Dutch satellite BeppoSax
discovered X-ray flashes \citep{hei01}, which may be closely related to long
gamma-ray bursts. Their non-thermal emissions are herein best 
accounted for by internal shocks in ultrarelativistic outflows 
in the fireball model proposed by \cite{ree92,ree94} -- 
with unknown inner engine producing these baryon-poor outflows. If gravitationally
powered, their compact size, based on short-time scale variabilities in the 
gamma-ray lightcurves, may involve the formation of black holes surrounded by
a high-mass torus \citep{pac91,woo93}.

At cosmological distances, the isotropic equivalent luminosities are on the 
order of $10^{51}$ erg s$^{-1}$ -- as bright as $10^{18}$ solar luminosities. 
These events, widely acclaimed as the ``Biggest Bang since the Big Bang", are now 
known to light up the Universe about once per minute. What, then, is their origin? 
The most powerful observational method to encyst the enigmatic energy source is 
through {\em calorimetry on all radiation channels}. This involves measuring the 
energy output in emissions in the electromagnetic spectrum, as well as in yet 
``unseen" emissions in neutrinos and gravitational waves. Perhaps thus shall we 
be in a position to determine their constitution, and decide whether their inner 
engine is {\em baryonic}, such as a neutron star, or {\em non-baryonic}: a black 
hole energized by spin. The exact solution of Roy P. \cite{ker63} -- see Table I 
for a parametrization -- shows an 
energy per unit mass which is anomalously large at high spin rates, exceeding 
that of a rapidly spinning neutron by an order of magnitude. By this potentially 
large and baryon-free energy reservoir, a Kerr black hole is a leading candidate 
for the inner engine of GRBs, at least those of long-durations in view of the 
following observational results.

The Italian-Dutch BeppoSax satellite, launched in 1996, dramatically changed the 
landscape of long GRBs with the discovery by E. Costa \citep{cos97} of X-ray 
afterglows in GRB970228 (confirmed in observations by the ASCA and ROSAT
satellites). These lower-energy emissions permit accurate localizations,
enabling follow-up by optical telescopes. Thus, J. van Paradijs \citep{par97} 
pointed the Isaac Newton Telescope and the William Herschel Telescope to further 
discover optical emissions associated with the same event. 
These lower-energy X-ray and optical emissions agree remarkably well with 
the expected decay of shocks in ultrarelativistic, baryon-poor outflows in 
the previously developed fireball model. Even lower-energy, radio-afterglows 
have been observed in some cases, including GRB970228 \citep{fra97}. The same
optical observations have now led to tens of GRBs with individually measured
redshifts, up to $z=6.29$ in GRB050904 detected by the recently
launched Swift satellite \citep{chi05,hai05}.

Optical follow-up identified the association of long GRBs with supernovae.
In particular, the unambiguous association of
GRB030329 with SN2003dh ($z=0.167,~D=800$Mpc) \citep{sta03} identified Type Ib/c 
supernovae as the parent population of long GRBs (Fig. 1). 
This observation confirms their association with core-collapse supernovae of 
massive stars, proposed by S. Woosley \citep{woo93,kat94,pac98}. Massive stars 
have characteristically short lifetimes, whereby GRB-supernovae track the 
cosmological event rate. This conclusion is quantitatively confirmed by 
consistent estimates of the true-to-observed GRB-SNe event rate of 450-500 deduced 
from two independent analysis, based on geometrical beaming factors \citep{fra01} 
and locking to the star-formation rate \citep{van03c}. These event rates 
correspond to a true cosmological event rate of about one per minute.

Core-collapse supernovae from massive stars are believed to produce
neutron stars or black holes. The former are perhaps best known from the rapidly 
spinning neutron star (a pulsar) in the Crab nebula, a remnant of what was 
probably a Type II supernova in 1054. The second is more difficult to ascertain 
with certainty, such as in the more recent Type II event SN1987 in
the Large Megellanic Cloud. At close proximity, its burst in MeV-neutrinos 
was detected by Kamiokanda and IMB (see \cite{bur87}) which provided direct 
evidence for the formation of matter at nuclear densities. The absence of 
a pulsar or a point-source of X-ray radiation in this case suggests
continued collapse -- probably into a stellar mass black hole.

Type II and Type Ib/c supernovae are both believed to represent the endpoint of
massive stars \citep{fil97,tur03}, and possibly so in binaries such as the 
Type II/Ib event SN1993J \citep{mau04}. This binary association suggests a hierarchy, 
wherein hydrogen-rich, envelope--retaining SNII are associated with wide binaries,
while hydrogen-poor, envelope--stripped SNIb and SNIc are associated with
increasingly compact binaries \citep{nom95,tur03}. By tidal coupling, the
primary star in the latter will rotate at approximately the orbital period
at the moment of core--collapse. With an evolved core \citep{bet03}, these
Type Ib/c events in particular are believed to produce a spinning black hole 
\citep{woo93,pac98,bro00,lee02}. However, the branching ratio of Type Ib/c
into GRB-supernovae is small,
\begin{eqnarray}
{\cal R}(\mbox{SNIb/c}\rightarrow \mbox{GRB}) 
  = \frac{ N(\mbox{GRB-SNe}) }{ N(\mbox{SNIb/c}) } \simeq (2-4)\times 10^{-3}
\label{EQN_branching}
\end{eqnarray}
as calculated from the true GRB-supernova event rate relative to the observed
event rates of supernovae of Type II and Type Ib/c. This ratio is remarkably
small, suggesting a higher-order down-selection process.

The identification of long GRBs with core-collapse supernovae brings into 
scope their potential gravitational-wave emissions. This may be due to a 
variety of mechanisms associated with rapidly rotating fluids. In particular, 
emissions will be produced by bar-mode instabilities producing fragmentation of 
orbiting matter prior to the formation of a compact object \citep{nak89,bon95} 
or in the formation of multiple compact objects \citep{dav02}, non-axisymmetries 
in accretion disks \citep{pap01,min02,kob03a,kob03b}, as well as after the formation 
into a rapidly spinning black hole.
The first may produce an initial ``splash" of gravitational radiation
\citep{ree74}, the second a bi-model spectrum containing high-frequency 
emissions produced by non-axisymmetric perturbations of the black hole and
low-frequency emissions from the disk, whereas the latter creates a 
long-duration burst of low-frequency gravitational radiation. 
Calculations this time-evolving spectrum of gravitational-radiation are
only beginning to be addressed by computational hydro- and magnetohydrodynamics,
during initial collapse \citep{ram98,fry99,mcf99,fry02,due04} and in the 
formation of a non-axisymmetric torus \citep{bro05}. 
Emissions from matter surrounding a stellar-mass black hole are in the 
frequency range of sensitivity of the 
gravitational-wave experiments LIGO \citep{abr92,bar99} and Virgo 
\citep{bra92,ace02,ale04}. Understanding their wave-forms can serve 
strategic search-and-detection algorithms, triggered by gamma-ray 
observations \citep{fin04} or by their associated supernova. The latter
is more common by the aforementioned true-to-observed event ratio of
GRB-supernovae.

In this review, we shall discuss a theory of GRBs from rotating black holes. 
Current phenomenology on GRB-supernovae poses a challenge to model
\begin{itemize}
\item The durations of long GRBs of tens of seconds 
\item The formation of an associated aspherical supernova
\item The launch of ultrarelativistic jets with
      beamed gamma-ray emissions 
\item A small branching ratio of Type Ib/c supernovae into GRBs of less than $0.5\%$
\end{itemize}
This phenomenology will be directly linked to the spin energy 
of the black hole. Quite generally, we are left with the task of modeling all 
possible radiation channels produced by the putative Kerr black hole, 
including gravitational radiation, MeV-neutrino emissions, and magnetic winds.

Just as lower-energy X-ray and optical afterglow emissions linked long GRBs to
supernovae, we expect that the detection of a contemporaneous burst in 
gravitational-radiation will lift the veil on their enigmatic inner--engine.
A precise determination of this link will enable the identification of a Kerr
black hole. The same might apply to Type II and Type Ib/c supernovae.

Understanding the energy source of long GRBs might tell us also about the constitution
of short GRBs. The first {\em faint} X-ray afterglows have been detected
of the short event GRB050509B by Swift \citep{geh05} and GRB050709 by HETE-II
\citep{vil05,fox05,hjo05}, further providing the low redshifts $z=0.225$ and 
$z=0.16$, respectively. The nature of host galaxies, preferably (older) 
elliptical galaxies for short bursts versus (young) star-forming galaxies 
for long bursts, supports the origin of short events in binary coalescence 
of black holes and neutron stars \citep{pir05}. After completion of this
review, \cite{kul05} outlined a very interesting prospect for long-lasting
supernova-like signatures to short bursts from the debris of a neutron star
binary coalescence with a similar partner or black hole.

Earlier, we explained the
dichotomy of short and long events in terms of hyper- and suspended accreting 
onto slowly and rapidly rotating black holes, respectively \citep{van01c}. 
Thus, we predicted similar X-ray afterglows 
with the property that those to short events are relatively faint: 
{\em the short burst is identical to the final moment of a long burst of gamma-rays}. 
Hyperaccreting, slowly rotating black holes can be produced through various channels.
Black-hole neutron star binaries are remnants of core-collapse events in binaries
(\cite{bet03}; unless formed by capture in dense stellar clusters). 
Any rapidly spinning black hole will be spun-down in the core-collapse process, 
possibly representing a prior GRB-supernova event (this review).
The merger of a neutron star onto a black hole \citep{pac91} could 
produce a state of hyperaccretion and produce a short gamma-ray burst. 
This scenario is consistent with the predominance of long over short events. 
Less clear is the potential for gamma-ray emissions from the merger of two 
neutron stars (e.g. \cite{zhu94,fab04}) which, however, should also produce
a slowly rotating black hole. Finally, slowly rotating black holes can also 
be produced in core-collapse of massive star in isolation or a wide binary.
Here as well, may detection of their gravitational-wave emissions provide 
a probe to differentiate amongst these different types of short bursts.

In \S2, we give a practical outlook on Kerr black holes and gravitational radiation.
In \S3, we review the formation and evolution (kick-velocities, growth and spin--down) 
of Kerr black holes in core--collapse supernovae. This includes the process of converting 
spin--energy into radiation and estimates on the lifetime of rapid spin.
In \S4, we calculate multipole mass-moments in a torus due to a hydrodynamical
instability. By asymptotic analysis, the energies in the various radiation channels 
are given in \S5.  An mechanism for launching ultrarelativistic jets from rotating
black holes by a small fraction of black hole spin is given in \S6. Specific
predictions for these ``flashes" as sources for the observatories LIGO and Virgo 
are given in \S7.

\section{Theoretical background}

Kerr black holes and gravitational radiation are two of the most dramatic
predictions of general relativity (other than cosmology). While evidence
for Kerr black holes, in particular in their role as powerful inner engines,
remains elusive, the quadrupole formula for gravitational radiation has
been confirmed observationally with great accuracy. 

\subsection{Energetics of Kerr black holes}

The Kerr metric describes an exact solution of a black hole spacetime with
nonzero angular momentum. It demonstrates the remarkable property of 
{\em frame-dragging}: the angular velocity $\omega$ of otherwise 
zero-angular momentum observers in spacetime outside the black hole. Frame-dragging
assumes a maximal and constant angular velocity $\Omega_H$ on the event horizon of 
the black hole, and decays with the cube of the distance to zero at large distances. 
This introduces an angular velocity $\Omega_H$ of the black hole, as well as 
{\em differential} frame-dragging by a non-zero gradient (whereby it is not gauge 
effect). The aforementioned energy reservoir in angular momentum satisfies, 
\begin{eqnarray}
E_{rot}=2M\sin^2(\lambda/4)
       =M_{irr}\left({\sqrt{1+(2M\Omega_H)^2}}-1\right),
\label{EQN_ROT}
\end{eqnarray}
where $\sin\lambda$ denotes the specific angular momentum of the black hole per unit 
mass. Here, we use geometric units with Newton's constant $G=1$ and velocity of light 
$c=1$. These properties give Kerr black holes the potential to react energetically to
their environment. They hereby have to potential of serving as universal sources of 
energy, distinct from any known baryonic object and with direct relevance to the 
phenomenology of GRBs. 

Rotating objects have a general tendency to radiate away their energy and angular 
momentum in an effort to reach a lower energy state. In the dynamics of rotating 
fluids, this is described by the well-known Rayleigh stability criterion. In many 
ways, black hole radiation processes are governed by the same principle. 
The second law of thermodynamics $dS\ge0$ for the entropy $S$ shows that the 
specific angular momentum of a Kerr black hole $(\Omega_H\le 1/2M$) increases 
with radiation:
\begin{eqnarray}
a_p\equiv \frac{-\delta J_H}{-\delta M} \ge \Omega_H^{-1} \ge 2M > M\ge a
\end{eqnarray}
for a black hole of mass $M$ and angular momentum $J_H$ which emits a particle with 
specific angular momentum $a_p$.  Generally, black holes in isolation are stable by 
exponential suppression of spontaneous emissions by canonical angular momentum barriers 
\citep{unr74,haw75,pre72,teu73,teu74}. Fortunately, magnetic fields can modulate and 
suppress these angular momentum barriers.

The earliest studies of active black holes focus on energy-extraction processes
in the ergosphere, i.e.: scattering of positive energy waves onto rotating black 
holes -- superradiant scattering of \cite{zel71}, \cite{pre72}, \cite{sta72} and 
\cite{bar72} -- as a continuous-wave analogue to the process of \cite{pen69}. A 
steady-state variation can be found in the process of spin--down through horizon 
Maxwell stresses, proposed \cite{ruf75} and extended to force-free magnetospheres 
by \cite{bla77}. Quite generally, therefore, magnetic fields are required as a 
mediating agent to stimulate luminosities which are of astrophysical interest.

Generally, stellar mass black holes produced in core-collapse of a massive star 
are parametrized by their mass, angular momentum and kick velocity $(M,J,K)$. 
The equilibrium electric charge of the black hole which arises in its lowest 
magnetic energy state, is relevant in sustaining adequate horizon magnetic flux 
especially at high spin--rates, but is insignificant relative to the spin-energy 
of the black hole.

\subsection{Linearized gravitational radiation}

Gravitational-wave emissions are predominantly due to low multipole mass-moments.
The quadrupole radiation formula according to the linearized theory of general
relativity has been given by \cite{pet63}
\begin{eqnarray}
L_{gw}=\frac{32}{5}\left(\omega {\cal M}\right)^{10/3} F(e), 
\label{EQN_PM}
\end{eqnarray}
where $F(e)$ is a function of the ellipticity $e$ of a binary with orbital
frequency $\omega$ and chirp mass ${\cal M}$. 

This linearized result has been observationally confirmed to within 0.1\% by 
decade-long observations of the Hulse-Taylor binary neutron star system PSR1913+16 
(with ellipticity $e=0.62$, \cite{hul75,tay94}). Its gravitational-wave luminosity 
of about about $0.15\%$ of a solar luminosity $L_\odot=4\times 10^{33}$ erg~s$^{-1}$
gives a binary lifetime of about 7.4 Gyr. Encouraged by this observational confirmation, 
we apply (\ref{EQN_PM}) to a non-axisymmetric matter surrounding a black hole. With 
orbital period $\omega\simeq M_H^{1/2}/R^{3/2}$, ${\cal M}\simeq M_H(\delta M_T/M_H)^{5/3}$ 
and ellipticity $e\simeq 0$, this predicts
\begin{eqnarray}
L_{gw}\simeq \frac{32}{5}\left(\frac{M_H}{R}\right)^5 \left(\frac{\delta M_T}{M_H}\right)^2
\label{EQN_MPT}
\end{eqnarray}
The quadrupole moment is due to a mass-inhomogeneity $\delta M_T$ in the torus.
A quadrupole moment appears spontaneously due to non-axisymmetric waves, such as 
a hydrodynamical or magnetohydrodynamical instability. A detailed description of 
radiation by all multipole mass-moments in a torus is given in \cite{bro05}.

These existing approaches to energy extraction from rotating black holes do not 
elucidate the ``loading problem:" the diversity in radiation channels and their
various energy outputs. This defines a novel challenge for Kerr black holes as 
inner engines to gamma-ray bursts. These events, perhaps including core-collapse
supernovae, pose the potential for a powerful link between Kerr black holes and 
gravitational radiation, as outlined below.

\section{Formation and evolution of black holes in core-collapse SNe}

While all SNIb/c might be producing black holes, only some are associated with 
GRB-supernovae in view of the observed small branching ratio (\ref{EQN_branching}).
Black holes produced in {\em aspherical} core-collapse receive typical kick 
velocities of a few hundred km s$^{-1}$ \citep{bek73}, measured relative to the 
center of mass of the progenitor star. Such objects inevitably escape from the 
central high-density region of the progenitor star, even before core-collapse is 
completed. A small sample of black holes receive low 
kick-velocities at birth. Remaining centered, these grow into rapidly spinning 
high-mass black holes by infall of a substantial fraction of the progenitor 
He-core mass (Fig. 2). This forms a starting point for the parametrization 
of rotating black holes as inner engines to GRB-supernovae.

The state of matter surrounding the newly formed black hole in core-collapse 
supernovae is unique, in attaining temperatures of in excess of 1MeV \citep{woo93} 
and masses of up to a few percent of the mass of the black hole \citep{van03b}. 
We further expect these accretion disks to be magnetized, representing a remnant 
magnetic field of the progenitor star and modified by a dynamo action in the disk.

The angular velocity $\Omega_H$ of the black hole easily exceeds that of surrounding 
matter, even at relatively low spin-energies. When a surrounding disk is largely 
unmagnetized, the black hole continues to accrete, enlarging it and spinning it up 
towards an extreme Kerr black hole along a {\em Bardeen trajectory} \citep{bar70}, 
shown in Fig. 2. Alternatively, the surrounding matter can be magnetized 
and, fairly generally, may contain an appreciable $m=0$ component of poloidal flux 
on average or, when time-variable, with non-zero standard deviation. Topologically,
this component represents a uniform magnetization, corresponding to two counter-oriented 
current rings \citep{van03b}. Under these conditions, a torus can form in a state of 
suspended accretion with angular velocity $\Omega_T<\Omega_H$. This introduces a channel
for {\em catalytic} conversion of spin--energy into various radiation channels 
\citep{van03b}, forcing the black hole to spin down until the angular velocities 
become similar $\Omega_H\simeq \Omega_T$.

\subsection{Catalyzing black--hole spin--energy}

A magnetized torus surrounding a black hole is expected to form in both 
black hole-neutron star coalescence and core--collapse of a massive star with 
very similar topological properties (Fig. 3). A magnetized star can be 
represented to leading order by a single current loop or equivalently a density of 
magnetic dipole moments. Stretching is around the black hole, as in tidal break-up, 
or by excision of the center of the star, as in core-collapse into a new black hole, 
leaves a magnetized annulus consisting of two counter-oriented current rings. Thus, 
core-collapse supernovae and binary black hole-neutron-star coalescence both give 
the same outcome: a black hole surrounded by a magnetized torus. In practice, the
resulting magnetic field will be modified by turbulence and, possibly, a dynamo.

We can look at the structure of the magnetic field of a torus by inspecting
its shape in a poloidal cross-section, and by comparison with pulsar magnetospheres.
In case of a pulsar, we see magnetic flux-surfaces of closed magnetic field-lines
that reach an outer light cylinder, while open magnetic field-lines extend
outwards to infinity. The same structure is found with regards to the torus'
outer face, when viewed in poloidal cross-section. Since the event horizon 
of the black hole is a null-surface, it has the same radiative boundary 
condition as asymptotic infinity \citep{bla77,tho86,oka92}, except for its 
finite surface area (and surface gravity).
It follows that the structure of magnetic field-lines found in a magnetized
neutron star also applies to the torus' inner face.
For the same reason that a spinning neutron star transfers angular 
momentum to infinity, the inner face of the torus gains angular momentum
from the black hole, whenever the latter spins more rapidly. This produces 
a {\em spin--connection by topological equivalence to pulsars}.

Angular momentum transport is by Alfv\'en waves emitted from the surface of
the torus, carrying positive angular momentum waves off the outer face to infinity
and negative angular momentum off the inner face into the black hole. 
In this process, the outer face becomes sub-Keplerian and the inner face becomes
super-Keplerian. On balance of the competing torques between these two faces
\citep{van03b}, this produces a state of suspended accretion \citep{van03b}.
The angular velocity of a Kerr black hole can reach $\Omega_H=1/2M$, which
far surpasses that of an accretion disk or torus. The black-hole spin--energy 
(\ref{EQN_ROT}) can be transferred to a torus with angular velocity $\Omega_T$ 
with efficiency
\begin{eqnarray}
\eta = \frac{\Omega_H}{\Omega_T}.
\label{EQN_ETA}
\end{eqnarray}
The task is now to quantify the various radiation channels provided by the torus.
We shall discuss these in the next sections.

The process of formation and spin--down of a Kerr black hole has been proposed 
to quantitatively model the supernovae by various groups \citep{bro00,bet03,van03b}. 
This approach represents a special case of aspherical, magneto-rotationally 
driven core-collapse supernovae, notably discussed for Type II supernovae 
by \cite{bis70,leb70,bis76,kun76,whe00,aki03}. 

\subsection{Durations and lifetime of rapid spin}

A given torus can support a finite poloidal magnetic field-energy.
This is due to an instability, produced by magnetic moment-magnetic 
moment self-interactions in the fluid.

Instability criteria can be derived for both tilt and buckling \citep{van03b}.
For a tilt instability between two counter-oriented current rings, representing
a uniformly magnetized torus, these interactions are described by a potential
\begin{eqnarray}
U_\mu(\theta)=-\mu B \cos\theta, 
\end{eqnarray}
where $\mu$ denotes the magnetic moment dipole moment of the inner ring, $B$ 
denotes the magnetic field produced by the outer ring, and $\theta$ denotes 
the angle between $\mu$ and $B$. Note that $U_\mu(\theta)$ has period $2\pi$, 
is maximal (minimal) when $\mu$ and $B$ are parallel (antiparallel). For the
buckling instability, the same interaction energies arise in the azimuthal 
partition of the torus into small current rings which, combined, are equivalent 
to the two counter-oriented current rings. 

The central potential well of the black hole provides a stabilizing contribution.
As the magnetic moment-magnetic moment interactions act primarily to introduce 
vertical displacements between two current rings (about their equilibrium 
configuration in the equatorial plane), we can focus on vertical displacements of 
fluid elements along surfaces of constant cylindrical radius $R$. This is distinct 
from motions of a rigid ring, whose fluid elements move on surfaces of constant 
spherical radius. In particular, the tilt of a current ring hereby changes the distance
to the central black hole according to $\rho = \sqrt{R^2 + z^2}\simeq R(1+z^2/2R^2)$. 
In the approximation of equal mass in the inner and outer face of the torus, 
simultaneous tilt of one ring upwards and the other ring downwards is associated with 
the potential energy (\cite{van03b}, corrected)
\begin{eqnarray}
U_g(\theta)\simeq -\frac{M_TM_H}{R}\left(1-\frac{1}{2}\tan^2(\theta/2)\right)
\end{eqnarray}
with $\tan(\theta/2)=z/R$ upon averaging over all segments of a ring. Note 
that $U_g(\theta)$ has period $\pi$ and is minimal at $\theta=0$. Similar
expressions hold for an azimuthal distribution of current rings.
Stability is accomplished provided that the total potential energy
$U(\theta)=U_\mu(\theta)+U_g(\theta)$ satisfies $U^{\prime\prime}(\theta)>0$.
We find the following poloidal magnetic field-to-kinetic energy ratios
\begin{eqnarray}
  \frac{{\cal E}_B}{{\cal E}_k}
  \simeq \left\{ 
  \begin{array}{rl}
  \frac{1}{6} & \mbox{tilt instability}\\
  \frac{1}{15}& \mbox{bucking instability}
\end{array}\right.
\label{EQN_EB1}
\end{eqnarray} 
in the approximation ${\cal E}_B=B^2R^3/6$, representing the poloidal magnetic 
field-energy of the inner torus magnetosphere in a characteristic volume $4\pi R^3/3$ 
and ${\cal E}_k = M_TM_H/2R$. We next discuss the physical parameters at this point 
of critical stability, which we interpret as a practical limit on the magnetic field 
energy that the torus can support.

For a pair of rings of radii $R_\pm$, $(R_+-R_-)/(R_+-R_-)=O(1)$, we have
$U_\mu(\theta)=(1/2)B^2 R^3 \cos\theta$, so that the point of critical
stability, $U^{\prime\prime}(\theta)=0$, gives for the critical magnetic field-strength
$B_c^2M^2_H=(1/4)(M_H/R)^4(M_T/M_H)$, or
\begin{eqnarray}
B_c = 10^{16}\mbox{G} \left(\frac{7M_\odot}{M_H}\right)
\left(\frac{6M_H}{R}\right)^2\left(\frac{M_T}{0.03M_H}\right)^{1/2},
\end{eqnarray}
with critical poloidal magnetic field-energy (\ref{EQN_EB1}). 

\begin{figure}
\plottwo{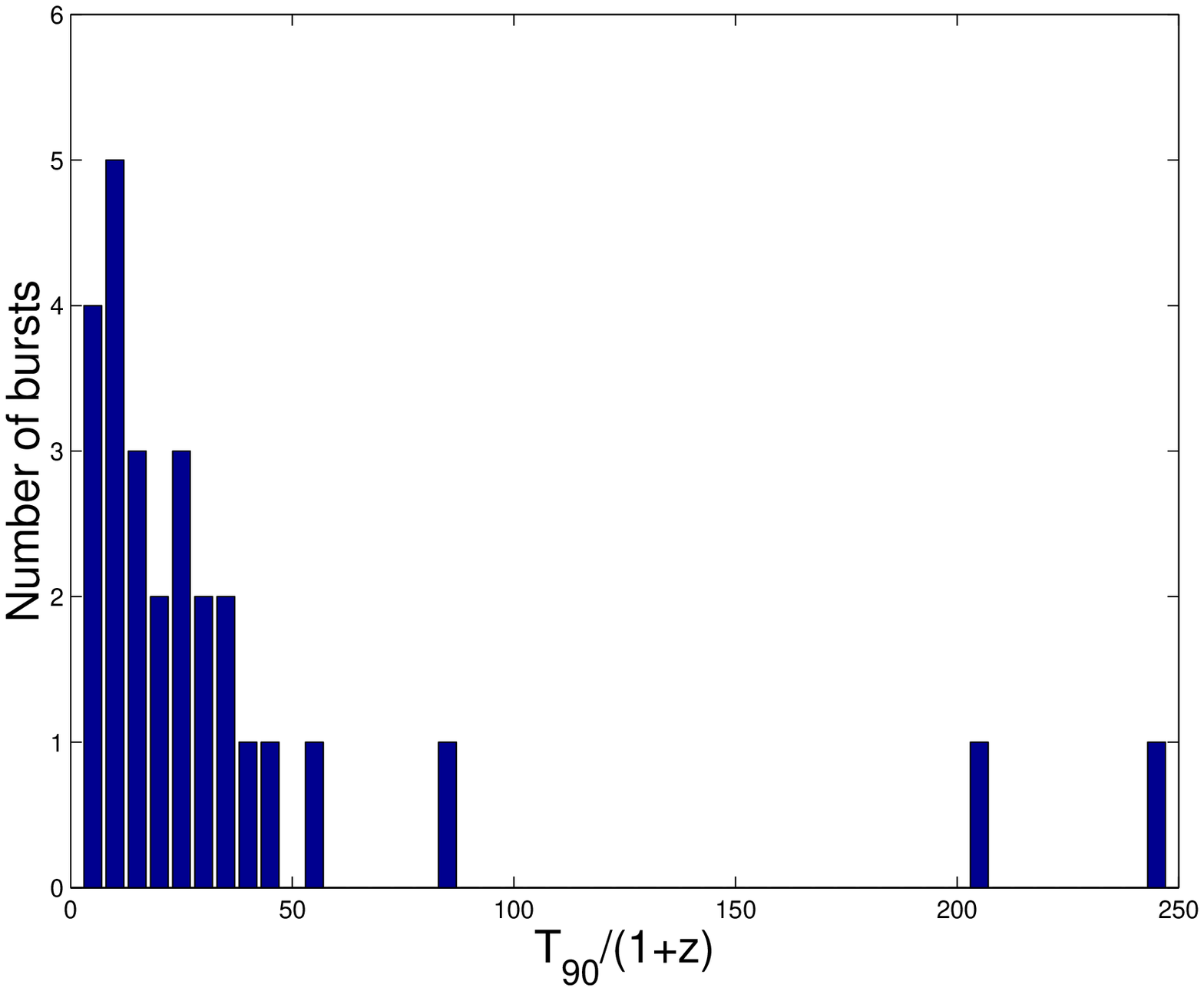}{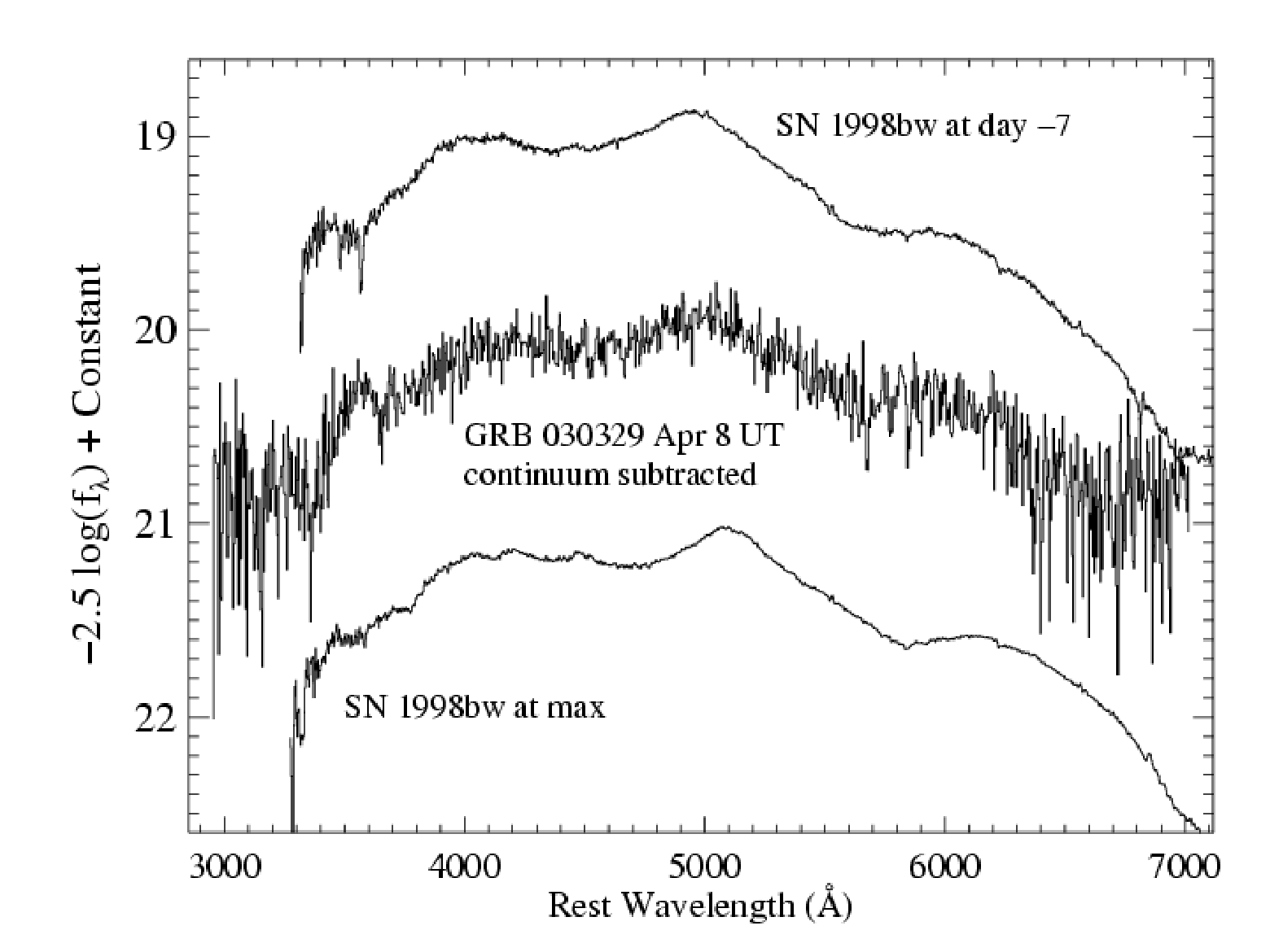}
\caption{}
\label{FIG_Y1}
\end{figure}

Rotating black holes with $\Omega_H\gg\Omega_T$ dissipate most of their spin--energy 
``unseen" in the event horizon. The lifetime of rapid spin is thus
\begin{eqnarray}
T_s \simeq \frac{E_{rot}}{T\dot{S}_H},~~~T\dot{S}_H\simeq \Omega_H^2A_\phi^2,
\end{eqnarray}
where $2\pi A_\phi$ denotes the horizon flux, taking into account that most
of the black-hole luminosity is incident onto the inner face of the torus.
We then have
\begin{eqnarray}
T_s \simeq 45\mbox{s}M_{H,7}\eta_{0.1}^{-8/3}\mu_{0.03}^{-1} E^{rot}_{0.5},
\label{EQN_TS}
\end{eqnarray}
where the subscripts denote normalization constants, i.e.: $M_{H,7}=M_H/M_\odot$, 
$\eta_{0.1}=\eta/0.1$, $\mu_{0.03}=M_T/0.03M_H$ and 
$E^{rot}_{0.5}={E_{rot}}/{0.5E_{rot,max}}$ corresponding to $\sin\lambda=0.8894$ 
in (\ref{EQN_ROT}). This estimate agrees well with the observed durations of tens 
of seconds of long GRBs shown in Fig. (\ref{FIG_Y1}) and statistics of the BATSE
catalogue \citep{kou93}.

\section{Formation of multipole mass-moments}

A torus tends to be unstable to a variety of symmetry-breaking instabilities.
This provides a spontaneous mechanism for the creation of multipole mass-moments.
Provided that the torus is not completely disrupted, such multipole mass-moments
ensure that the torus is luminous in gravitational radiation, when its mass reaches
a few percent of that of the black hole. These instabilities can take the form of
hydrodynamic and magnetohydrodynamic instabilities.

\cite{pap84} describe a hydrodynamic buckling instability in infinitesimally slender tori, 
due to a coupling of surface waves on a super-Keplerian inner face and a sub-Keplerian 
outer face. This theory can be extended to tori of arbitrary slenderness to be of 
practical interest (below). In a recent numerical study, the magnetic moment-magnetic 
moment self-interactions of a uniformly magnetized torus are also found to produce 
instabilities -- buckling instabilities in the poloidal plane \citep{bro05}. This 
fully nonlinear numerical study, though one-dimensional, recovers our heuristic 
analytical bound on the maximal poloidal magnetic field energy-to-kinetic energy 
in the torus (\ref{EQN_EB1}). It confirms that the dominant emission channel in
gravitational radiation is through the quadrupole mass-moment. These are but two 
example calculations on the more general challenge of computing gravitational-wave
spectra of magnetized tori in suspended accretion. Below, we discuss the
hydrodynamic buckling instability for wide tori. 

The dynamical stability of a torus with sub-Keplerian outer face and super-Keplerian
inner face can be studied in the limit of an inviscid incompressible fluid with
Newtonian angular velocity \citep{pap84,gol86}
\begin{eqnarray}
\Omega(r) = \Omega_a\left(\frac{a}{r}\right)^q ~~~(3/2\le q \le 2).
\label{EQN_OME}
\end{eqnarray}
Here, $q$ denotes the rotation index which is bounded between the Keplerian
value $q=3/2$ and Rayleigh's stability limit $q=2$. 

Irrotational perturbations to the underlying flow (vortical if $q\ne 2$) remain
irrotational by Kelvin's theorem. In studying their stability we consider the
harmonic velocity potential in cylindrical coordinates $(r,\theta,z)$
\begin{eqnarray}
\phi = \Sigma_n a_n(r,\theta,z) z^n,~~~\Delta \phi=0.
\label{EQN_PHI}
\end{eqnarray}
The equations of motion can be expressed in a local Cartesian frame 
$(x,y,z)$ with Newtonian angular velocity $\Omega_a = M^{1/2}a^{-3/2}$,
equal to the angular velocity of the torus at its major radius $r=a$
about a central mass $M$. These Cartesian coordinates are related to
cylindrical coordinates $(r,\theta)$ according to 
$x=r-a$, $\partial_x=\partial_r$ and $\partial_y=r^{-1}\partial_\theta$.
Together with zero-enthalpy boundary conditions on the free inner and
outer surface of the torus, we obtain a complete problem for linearized
stability analysis. This problem can be solved semi-analytically, by
searching numerical for points of change in stability \citep{kel87} of
harmonic perturbations $e^{im\theta - i\omega^\prime t}$ of infinitesimal
amplitude at frequency $\omega^\prime$ as seen in a corotating frame
at $r=a$. Fig. 4 shows the numerical stability diagram. 

In general, we encounter for each $m$ a critical rotation index $q=q(b/a,m)$ 
which depends on the minor-to-major radius $b/a$ of the torus. These
curves can be found using numerical continuation methods \citep{kel87}.
The stability diagram for the rotation index is shown in Fig. 4.
In particular, we mention the critical values 
\begin{eqnarray}
b/a=0.7506, 0.3260, 0.2037, 0.1473, 0.1152, \cdots, 0.56/m
\label{EQN_BAM}
\end{eqnarray}
for the various $m$-modes at the Rayleigh stability line $q=2$.

\section{Radiation-energies by a non-axisymmetric torus} 

Rotating black holes in suspended accretion dissipate most of their
rotational energy (\ref{EQN_ROT}) ``unseen" in their event horizon, 
while a major fraction (\ref{EQN_ETA}) is incident into the inner
face of the surrounding torus. This lasts for the lifetime of rapid spin 
(\ref{EQN_TS}), whose tens of seconds represents a secular timescale 
relative to the millisecond period of the orbital motion of the torus. 
This reduces the problem of calculating the radiation output from the 
torus to algebraic equations of balance in energy and angular momentum flux, 
taking into account the channels of gravitational radiation, MeV-neutrino 
emissions and magnetic winds. The equations of suspended accretion are
\begin{eqnarray}
\begin{array}{rl}
\tau_+ & = \tau_- + \tau_{gw}\\
\Omega_+\tau_+ & = \Omega_-\tau_- + \Omega_T \tau_{gw} + P_\nu,
\end{array}
\label{EQN_SAS}
\end{eqnarray}
where ($\tau_\pm,\Omega_\pm,\Omega_T=(\Omega_++\Omega_-)/2)$ denote the 
torques on and angular velocities of the inner and outer face of the torus due 
to surface Maxwell stresses as those on a pulsar, $\tau_{gw}$ denotes the 
torque on the torus due to the emission of gravitational radiation with
luminosity $L_{gw}=\Omega_T\tau_{gw}$ and $P_\nu$ the power in MeV-neutrino 
emissions due to dissipation. (The results show temperatures of a few
MeV.) These equations are closed by a constitutive relation for the
dissipation process. We set out by attributing dissipative heating to 
magnetohydrodynamical stresses. This introduces an overall scaling with
the magnetic field-energy $E_B$. The resulting total energy emissions,
$E_{gw}$, $E_w$ and $E_\nu$, produced over the lifetime of rapid spin
of the black hole (\ref{EQN_TS}), thus become {\em independent} of $E_B$
and reduce to specific fractions of $E_{rot}$.

When the torus is sufficiently slender, $m=2$ modes (two lumps swirling around the black 
hole) develop. These radiate at essentially twice the orbital frequency of the torus, 
when the minor-to-major radius is less than 0.3260 by (\ref{EQN_BAM}).

By an asymptotic analysis of (\ref{EQN_SAS}), the gravitational-wave emissions
satisfy \citep{van03b}
\begin{eqnarray}
   E_{gw} \simeq 2\times 10^{53}~\eta_{0.1} M_{H,7}~E^{rot}_{0.5}~\mbox{erg},
~~~f_{gw}\simeq 500\eta_{0.1}M_{H,7}^{-1}~\mbox{Hz}.
\label{EQN_EGW}
\end{eqnarray}
Thus, an ``unseen" energy output (\ref{EQN_EGW}) surpasses the true energy 
$E_\gamma\simeq 3\times 10^{50}$erg in gamma-rays \citep{fra01} by several 
orders of magnitude. 

The associated output in magnetic winds contains an additional factor $\eta$, i.e.,
\begin{eqnarray}
E_w \simeq 2\times 10^{52} ~\eta_{0.1}^2 M_{H,7} E^{rot}_{0.5} ~\mbox{erg}.
\label{EQN_EW}
\end{eqnarray}
This baryon-rich wind provides a powerful agent towards collimation of any
outflows from the black hole \citep{lev00}, as well as a source of neutrinos 
for pick-up by the same \citep{lev03}. It is otherwise largely incident onto the 
remnant stellar envelope {\em from within}, which serves as energetic input to 
accompanying supernova ejecta. We estimate these kinetic energies to be 
\begin{eqnarray}
E_{SN} \simeq 1\times 10^{51}~\beta_{0.1}M_{H,7}\eta_{0.1}^2 E^{rot}_{0.5}~\mbox{~erg} 
\label{EQN_SN}
\end{eqnarray}
with $\beta=v_{ej}/c$ denoting the velocity $v_{ej}$ of supernova ejecta relative 
to the velocity of light $c$. In the expected aspherical geometry, $\beta=0.1\beta_{0.1}$ 
refers to the mass-average taken over all angles. The canonical value $\beta=0.1$ refers 
to the observed value in GRB011211 \citep{ree02}. Eventually, the expanding remnant 
envelope becomes optically thin, which permits the appearance of X-ray line-emissions 
excited by the underlying continuum emission $E_\gamma$ by dissipating $E_w$. 
The estimate (\ref{EQN_SN}) is in remarkable agreement with the kinetic energy 
$2\times 10^{51}$ erg of the aspherical supernova SN1998bw associated with GRB980425 
\citep{hoe99}. The energy output in MeV-neutrinos is intermediate, in being smaller 
than $E_{gw}$ by an additional factor given by the slenderness parameter 
$\delta = qb/2R$, where $b/R$ denotes the ratio of minor-to-major radius of the torus 
and $q$ its rotation index (\ref{EQN_OME}). Thus, we have
\begin{eqnarray}
E_\nu \simeq 1\times 10^{53} ~\eta_{0.1} 
             \delta_{0.30} M_{H,7}E^{rot}_{0.5} ~\mbox{erg},
\end{eqnarray}
where $\delta=0.30\delta_{0.30}$. At the associated dissipation rate, the torus develops
a temperature of a few MeV which stimulates the production of baryon-rich winds
\citep{van03b}.

\section{Launching an ultrarelativistic jet by differential frame-dragging}

Frame-dragging induced by angular momentum extends through the environment of the
black hole and includes the spin--axis. The Kerr metric provides an exact description 
of this gravitational induction process by the Riemann tensor \citep{cha83}. In turn, 
the Riemann tensor couples to spin \citep{pap51a,pap51b,pir56,mis74,tho86}. 
Specific angular momentum (angular momentum per unit mass) represents a rate of 
change of surface area per unit of time, while the Riemann tensor is of dimension 
cm$^{-2}$ (in geometrical units). Therefore, curvature--spin coupling produces a 
force (dimensionless in geometrical units), whereby test particles with spin follow 
non-geodesic trajectories \citep{pir56}. In practical terms, the latter holds promise
as a mechanism for {\em linear acceleration}.
By dimensional analysis once more, the gravitational potential for spin--aligned 
interactions satisfy 
\begin{eqnarray}
E=\omega J
\label{EQN_EO}
\end{eqnarray}
where $\omega$ denotes the local frame-dragging angular velocity produced
by the black hole (or any other spinning object) and $J$ is the angular 
momentum of the particle at hand. Spinning bodies therefore couple to spinning
bodies \citep{oco72}. Thus, (\ref{EQN_EO}) defines a mechanism for 
accelerating baryon-poor ejecta to ultrarelativistic velocities, provided
$J$ is large. 

The angular momentum $J$ of charged particles in strong magnetic fields, 
confined to individual magnetic flux-surfaces in radiative Landau states,
is macroscopic in the form of orbital angular momentum 
\begin{eqnarray}
J=eA_\phi.
\label{EQN_EO1}
\end{eqnarray}
Here $e$ denotes the elementary charge and $2\pi A_\phi$ denotes the enclosed 
magnetic flux, where $A_a$ denotes the electromagnetic vector potential of an
open magnetic flux-tube along the spin--axis of the black hole.
Geometrically, the specific angular momentum represents the rate of change 
of surface area traced out by the orbital motion of the charged particle,
by ``helical motion" as seen in four-dimensional spacetime.
 
Combined, (\ref{EQN_EO}) and (\ref{EQN_EO1}) describe a powerful mechanism for linear
acceleration of baryon-poor matter. It radically differs from the common view, that 
black-hole energetic processes are limited exclusively to the action of frame-dragging 
in the ergosphere. 

It is instructive to derive (\ref{EQN_EO}) by specializing to the Kerr metric
on the spin--axis of the black hole. To this end, we may consider one-half the 
difference in potential energy of particles with angular momenta $\pm J$ suspended 
in a gravitational field about a rotating object. In the first case, we note that the
Riemann tensor $R_{ijmn}$ expressed relative to tetrad elements associated with
Boyer-Lindquist coordinates \citep{cha83} gives rise to a linear force
\begin{eqnarray}
F_2 = J R_{3120} = -\partial_2 \omega J
\end{eqnarray}
which can be integrated along the spin--axis to give
\begin{eqnarray}
E=\int_r^\infty F_2 ds = \omega J.
\end{eqnarray}
In the second case, we merely assume a metric $g_{ab}$ with
time-like and azimuthal Killing vectors and consider 
two particles with velocity four-vectors $u^b$ according to angular momenta 
$J_\pm = g_{\phi\phi} u^t (\Omega_\pm - \omega)$,
\begin{eqnarray}
J_\pm  = \pm g_{\phi\phi} u^t \sqrt{\omega^2 - (g_{tt} + (u^t)^{-2}/g_{\phi\phi}} = \pm J,
\end{eqnarray}
which shows that $u^t$ is the same for both particles. The total energy of
the particles $E_\pm = (u^t)^{-1} + \Omega_\pm$ gives rise to
\begin{eqnarray}
E=\frac{1}{2}\left( E_+ - E_- \right) = \omega J.
\end{eqnarray}

To apply (\ref{EQN_EO}-\ref{EQN_EO1}) to gamma-ray bursts from rotating black
holes, consider a perfectly conducting blob of charged particles in a magnetic 
flux-tube subtended at a finite half-opening angle $\theta_H$ on the event horizon of the black
hole and along its spin--axis. In the approximation of electrostatic equilibrium, it 
assumes a rigid rotation \citep{tho86} described by an angular velocity $\Omega_b$.
(By Faraday induction, differential rotation introduce potential differences along 
magnetic field-lines. Some differential rotation is inevitable over long length scales 
and, possibly, in the formation of gaps). In the frame of zero-angular 
momentum observers, the equilibrium charge-density assumes the value of \cite{gol69},
as viewed by zero-angular momentum observers. Thus, given a number $N(s)$ of charged 
particles per unit scale height $s$ in the flux-tube, 
\begin{eqnarray}
N(s)=(\Omega_b-\omega)A_\phi/e. 
\end{eqnarray}
A pair of blobs of scale height $h=h_MM_H$ in both directions along the spin--axis 
of the black hole each hereby receive a potential energy
\begin{eqnarray}
E_{b} = \omega JNh = 10^{47} B_{15}h^3_M H \mbox{~erg},
\label{EQN_EB}
\end{eqnarray}
where $B=B_{15}\times 10^{15}$G and $H=4\hat{\omega}\left(\hat{\Omega}_b
-\hat{\omega}\right)$ is a quantity of order unity, expressed in terms 
of the normalised angular velocities $\hat{\omega}=\omega/\Omega_H$
and $\hat{\Omega}_b = \Omega_b/\Omega_H$.

Electrons and positrons in superstrong magnetic fields are essentially
massless. The ejection of a pair of blobs with energy (\ref{EQN_EB}) 
thus takes place in a light-crossing time of about 0.3ms of, e.g.,
seven solar mass black hole of linear dimension $10^7$cm. This corresponds 
to in an instantaneous luminosity of about $3\times 10^{50}$erg/s. This 
produces a total kinetic energy output of up to $10^{52}$ erg in tens of seconds
in long bursts. A fraction hereof which will be dissipated in gamma-rays and 
lower-energy afterglow emissions in the internal-external shock model for GRBs 
\citep{ree92,ree94}.

Similar results obtain by considering the luminosity in a steady-state limit,
by considering a horizon half-opening angle $\theta_H\simeq {M_H}/{R}$,
set by the poloidal curvature $M_H/R$ of the magnetic field-lines. We note
the paradoxical {\em small} energy output of GRB-afterglow emissions, when 
viewed relative to the total black-hole spin--energy (\ref{EQN_ROT}). The
luminosity in the jet scales with the square of the enclosed magnetic flux,
while the latter scales with the enclosed surface area, and hence the square 
of the half-opening angle $\theta_H$. Thus, we encounter a geometrical 
scale-factor, which creates a jet luminosity
\begin{eqnarray}
L_j\propto \theta_H^4.
\end{eqnarray} 
Even when $\theta_H$ is not small, e.g., about $10$ degrees, this scaling 
creates a small parameter. Integrated over time, the energy output in gamma-rays
satisfies \citep{van03b}
\begin{eqnarray}
E_\gamma\simeq 1\times 10^{50} \epsilon_{0.3}\eta_{0.1}^{8/3}~E^{rot}_{0.5}~\mbox{erg},
\end{eqnarray}
consistent with the observed true energy output $3\times 10^{50}$ erg in gamma-rays 
of long bursts \citep{fra01}, where $\eta=0.1\eta_{0.1}$ and 
$\epsilon=0.3\epsilon_{0.3}$ denotes the efficiency of converting kinetic energy to 
gamma-rays.

In our unification scheme, the durations of short and long bursts are attributed
to different spin-down times of, respectively, a slowly and rapidly spinning
black hole according to (\ref{EQN_TS}). The total energy output in gamma-rays
of short GRBs is hereby significantly smaller than that of long GRBs, due to
both shorter durations and smaller luminosities. In this light, the recent
X-ray afterglow detections to short bursts, very similar but fainter than
their counterparts to long bursts, are encouraging.

\section{Long-duration bursts in gravitational-waves}

The output in gravitational radiation (\ref{EQN_EGW}) is in the
range of sensitivity of the broad band detectors LIGO  
and Virgo shown in Fig. 5, as 
well as GEO \citep{dan95,wil02} and TAMA \citep{and02}. Note that the
match is better for higher-mass black holes, in view of their emissions
at lower frequencies (towards the minimum in the detector noise curve)
and their larger energy output.

Matched filtering gives a theoretical upper bound on the signal-to-noise 
ratio in the detection of long bursts in gravitational radiation from
GRB-supernovae. In practice, the frequency will be unsteady at least on 
an intermittent timescale associated with the evolution of the torus, 
e.g., due to mass-loss in winds and possibly mass-gain by accretion from 
additional matter falling in. For this reason, a Time-Frequency Trajectory 
method which correlates the coefficients of a Fourier transformation
over subsequent windows of durations of seconds might apply. 
The ultimate signal-to-noise ratio will therefore
be intermediate between that obtained through 
correlation -- a second-order procedure -- and matched filtering -- a
first-order procedure. The results shown in Fig. 5
show the maximal attainable signal-to-noise ratio (by matched
filtering) for sources at a fiducial distance of 100Mpc. This distance 
corresponds to an event rate of one per year.

GRB-supernovae are an astrophysical source population locked to the star-formation rate.
We can calculate their contribution to the stochastic background in gravitational 
radiation accordingly given their band-limited signals, assuming
$B=\Delta f/f_e$ of around 10\%, where $f_e$ denotes the average
gravitational-wave frequency in the comoving frame. In what follows,
we the following scaling relations are applied,
\begin{eqnarray}
E_{gw} = E_0 M_H/M_0,~~~f_e = f_0 M_0/M_H
\end{eqnarray}
where $M_0=7M_\odot$, $E_0=0.203M_\odot \eta_{0.1}$ and
$f_0=455$Hz$\eta_{0.1}$, assuming maximal spin--rates 
($E_{rot}=E_{rot,max}$). For non-extremal black holes, a
commensurate reduction factor in energy output can be inserted.
This factor carries through proportionally to the final results, 
whence it is not taken into account explicitly.

Summation over a uniform distribution of black hole masses,
e.g., $M_H=(4--14)\times M_\odot$, and assuming that the
black hole mass and $\eta$ in (\ref{EQN_ETA}), are uncorrelated,
the expected spectral energy-density satisfies \citep{van04e}
\begin{eqnarray}
<\epsilon_B^\prime(f)> = 1.08 \times 10^{-18} \hat{f}_B(x)
\mbox{~erg~cm}^{-3}~\mbox{Hz}^{-1}
\end{eqnarray}
where $\hat{f}_B(x) = f_B(x)/$max$f_B(\cdot)$ is a normalized
frequency distribution. The associated dimensionless amplitude
$\sqrt{S_B(f)}=\sqrt{2G/\pi c^3} f^{-1} \tilde{F}_B^{1/2}(f)$, where
$\tilde{F}_B=c\epsilon_B^\prime$ and $G$ denotes Newton's constant
satisfies
\begin{eqnarray}
\sqrt{S_B(f)}=7.41\times 10^{-26}\eta_{0.1}^{-1} \hat{f}_S^{1/2}(x)
\mbox{~Hz}^{-1/2}
\end{eqnarray}
where $\hat{f}_S(x) = f_S(x)/$max$f_S(\cdot)$, $f_S(x)=f_B(x)/x^2$,
Likewise, we have for the spectral closure density $\Omega_B(f)
=f \tilde{F}_B(f)/\rho_cc^3$ relative to the closure density
$\rho_c = 3H_0^2/8\pi G$
\begin{eqnarray}
\tilde{\Omega}_B(f)= 1.60\times 10^{-8} \eta_{0.1} \hat{f}_\Omega(x),
\end{eqnarray}
where $\hat{f}_\Omega(x) = f_\Omega(x)/$max$f_\Omega(\cdot)$,
$f_\Omega(x)=xf_B(x)$ and $H_0$ denotes the Hubble constant. 

These cosmological results show a simple scaling relation for the
extremal value of the spectral closure density in its dependency
on the model parameter $\eta$. The location of the maximum
scales with $f_0$ in view of $x=f/f_0$. The spectral closure
density thus becomes completely determined by the shape of
the function representing the star-formation rate, the
fractional GRB-supernova rate thereof, $\eta$, and the black-hole
mass distribution. Fig. 5 shows the various
distributions. The extremal value of $\Omega_B(f)$ is in the
neighborhood of the location of maximal sensitivity of LIGO and
Virgo. It would be of interest to search for this contribution
to the stochastic background in gravitational waves by
correlation in the spectral domain, following Fourier transformation
over series of sub-windows on intermediate timescales of seconds.

\section{Conclusions}

The sixties saw two independent discoveries: the first GRB670702 by the Vela 
satellite and the exact solution of rotating black holes by Roy P. Kerr. 
Through observational campaigns with BATSE, BeppoSax, the Interplanetary Network 
(IPN), HETE-II and now Swift, we have come to understand the phenomenology of GRBs. 
Long bursts are association with supernovae, representing a rare and extraordinary 
powerful cosmological transient, taking place about once a minute and reaching the
earliest epochs in the Universe. Independently, through the works of 
\cite{pen69,ruf75,bla77} and others, we have come to understand the potential
significance of Kerr black holes as compact, baryon-free energy sources with
certain universal properties. The applications to high-energy astrophysics should
be enormous (e.g. \cite{lev04}), of supermassive black holes in active galactic nuclei 
including our own galaxy \citep{por04}, and of stellar mass black holes in microquasars 
\citep{mir94} and, possibly, gamma-ray bursts.

While the formation-process of supermassive black holes remains inconclusive, the 
birthplace of stellar mass black holes is most probably core-collapse supernovae of 
massive stars. Evidence for Kerr black holes as the energy source to high-energy 
astrophysical processes remains elusive, however. Recent measurements on frame-dragging 
by X-ray spectroscopy, typically during inactive states of the black hole, are 
encouraging in this respect (A.C. Fabian, these proceedings). 

Specically, we propose that GRB-supernovae are powered by rapidly rotating
black holes, wherein (1) the durations of long GRBs of tens of seconds
are identified with the lifetime of rapid spin of the black hole in a state
of suspended accretion, (2) an accompanying supernova is radiation-driven by 
magnetic winds from a torus in suspended accretion, (3) ultrarelativistic 
outflows are launched by gravitational spin-orbit coupling with charged 
particles along open magnetic field-lines, and (4) a small branching ratio 
of Type Ib/c supernovae into GRBs is attributed to the small probability of 
producing black holes with small kick velocities.

Modeling short GRBs from slowly rotating black holes, we predicted X-ray afterglows
very similar but weaker (in total energies) than those of long bursts. The recent
discovery of faint X-ray afterglows to GRB050509B and GRB050709 fit well within this
scheme. 
 
Gamma-ray bursts present a potentially powerful link between rotating black holes and 
gravitational radiation. Strategic searches for their chirps in binary
coalescence of neutron stars and black holes, or long-duration bursts in gravitational
radiation during radiative spin-down of a high-mass black hole, can be pursued by
the advanced detectors LIGO and Virgo. Strategic searches are preferrably pursued
in combination with upcoming optical-radio supernova surveys, e.g., Pan-Starrs in 
Hawaii \citep{kud03} in combination with the Low Frequency Array \citep{lofar}, see 
also \cite{gal05}. In light of the proposed supernova-like signatures from the debris
of a neutron star \citep{kul05}, these strategies might apply to both short and long
bursts.

LIGO and Virgo promise to bring together a serendipitous discovery and general
relativity, providing a unique method to identify Kerr black holes as the most 
luminous objects in the Universe.

{\bf Acknowledgment.}
The author thanks A. Levinson, R.P. Kerr, R. Preece, and David Wiltshire 
for constructive comments. This research 
is supported by the LIGO Observatories, constructed by Caltech and MIT
with funding from NSF under cooperative agreement PHY 9210038.
The LIGO Laboratory operates under cooperative agreement PHY-0107417. 
This paper has been assigned LIGO document number LIGO-P040013-00-R.

\newpage
\centerline{\bf Table Captions}

{\bf Table I.}
Trigonometric parametrization of a Kerr black hole. Here, $M$ denotes
the mass of the black hole, $a=J_H/M$ denotes the specific angular momentum,
$E_{rot}$ the rotational energy and $M_{irr}$ denotes the irreducible mass.

\mbox{}\\
\centerline{\bf Figure Captions}

{\bf Fig. 1} 
$(Left)$ A histogram of redshift-corrected distributions of 27
long bursts with individually determined redshifts from their afterglow
emissions. It shows durations $T_{90}/(1+z)$ of tens of seconds at a mean
redshift distance $<z>=1.25$, indicative of their cosmological distances.
$(Right)$ The optical spectrum of the Type Ic SN2003dh associated with
GRB030329 is remarkably similar to that of the Type Ic SN1998bw of GRB980425
one week before maximum. GRB030329 displayed a gamma-ray luminosity of about
$10^{-1}$ times typical at a distance of $z=0.167$ ($D\simeq 800$Mpc), whereas
GRB980425 was observed at anomalously low gamma-ray luminosity
$(10^{-4}$ times typical) in the local universe ($z=0.008, D\simeq 37$Mpc).
At the same time, their supernovae were very luminous with inferred 
$^{56}$Ni ejecta of about 0.5$M_\odot$. 
(Reprinted from \cite{van02,sta03}\copyright The American Astronomical Society.)

{\bf Fig. 2} 
($Left$) Black holes with small kick velocities remain centered in core-collapse of a uniformly rotating
massive star. Shown is the accumulated specific angular momentum of the central object (arbitrary units) 
versus dimensionless orbital period $1/\beta$. Arrows indicate the evolution as a function
of time. Kerr black holes exist {\em inside} the outer curve (diamonds). A black hole
forms in a first-order transition following the formation and collapse of a torus. This 
produces a short burst in gravitational radiation. When centered, the black hole surges 
to a high-mass object by direct infall of matter with relatively low specific angular 
momentum, up to the inner continuous curve (ISCO). At this point, the black hole either 
spins up by continuing accretion or spins down radiatively against gravitational radiation 
emitted by a surrounding non-axisymmetric torus. In this state, the black hole creates a 
baryon poor jet as input to GRB-afterglow emissions. This continues until the angular 
velocity of the black hole equals that of the torus (dot-dashed line). This scenario fails 
for black holes with typical kick velocities with inevitable escape from the high-density 
core, prohibiting the formation of a high mass black hole surrounded by a high-mass torus.
The probability of small kick velocities defines the branching ratio of Type Ib/c 
supernovae into long GRBs. ($Right$) The black hole mass $M$ and rotational energy 
$E_{rot}$ of formed after surge in case of small kick velocities, expressed relative to 
the mass $M_{He}$ of the progenitor He-star. The results are shown in cylindrical geometry 
(continuous) and spherical geometric (dashed). Note the broad distribution of high-mass 
black holes with large rotational energies of $5-10\%$ (spherical to cylindrical) of
$M_{He}c^2$. (Reprinted from \cite{van04b}\copyright{}The American Astronomical Society.)

{\bf Fig. 3}
A uniformly magnetized torus represented by two counter-oriented
current rings around a black hole (C) forms out of both core-collapse (A1-B1) 
in massive stars and tidal break-up (A2-B2) in black hole-neutron star 
coalescence, followed by a single reconnection event (B2-C). 
(Reprinted from \cite{van03b}\copyright{}The American Astronomical Society.)

{\bf Fig. 4} 
The stability diagram showing the neutral stability
curves for non-axisymmetric buckling modes in a torus of an inviscid
incompressible fluid of arbitrary minor-to-major radius $b/a$. Curves
of critical rotation index $q_c$ are labeled with azimuthal quantum
number $m=1,2,\cdots$, where instability sets in above and stability
sets in below. (Reprinted from \cite{van02}\copyright{}The American Astrophysical 
Society.)

{\bf Fig. 5} 
($Left$) GRB-supernovae from rotating black holes predict a contemporaneous
long-duration burst in gravitational radiation within the range of sensitivity
of upcoming gravitational-wave detectors LIGO and Virgo within a distance of
about 100Mpc. The corresponding event rate is about one per year. The ``blue
bar" denotes the distribution of dimensionless strain amplitudes for a
distribution of sources corresponding to a range of black hole masses and
efficiency factors, assuming matched filtering. In practice, the sensitivity
will depend less on using time frequency trajectory methods with correlations
in the spectral domain. $(Right)$ The cosmological distribution of GRB-supernovae
is locked to the star-formation rate $N(z)$. This enables the calculation of the
expected contribution to the stochastic background in gravitational radiation,
here shown in terms of the spectral energy-density $\epsilon_B^\prime(f)$,
the strain amplitude $S_B^{1/2}(f)$ and the spectral closure density
$\Omega_B(f)$. The results are calculated for uniform mass-distributions
of the black hole, $M_H=(4-14)\times M_\odot$ (top curves) and $M_H=(5-8)\times M_\odot$
(lower curves) with $\eta=0.1$ (solid curves) and $\eta=0.2$ (dashed curves).
The extremal value of $\Omega_B(f)$ is in the neighborhood of maximal 
sensitivity of LIGO and Virgo. (Reprinted from \cite{van04e}\copyright2004
American Physical Society.)

\newpage
\begin{table}
\begin{center}
\caption{}
\mbox{~}\\
\begin{tabular}[t]{ll@{\quad}c@{\quad}c@{\quad}}
 {\sc Symbol} &  {\sc Expression} & {\sc Comment} \\
 \hline\\
 $\lambda$    &  $\sin\lambda = a/M$  & \\
 $\Omega_H$   &  $\tan(\lambda/2)/2M$ & \\
 $J_H$        &  $M^2\sin\lambda$  & \\
 $E_{rot}$    &  $2M\sin^2(\lambda/4)$& $\le 0.29 M$\\
 $M_{irr}$    &  $M\cos(\lambda/2)$   & $\ge 0.71 M$\\
 \mbox{}\\\hline
\end{tabular}
\end{center}
\end{table}

\begin{figure}
\plottwo{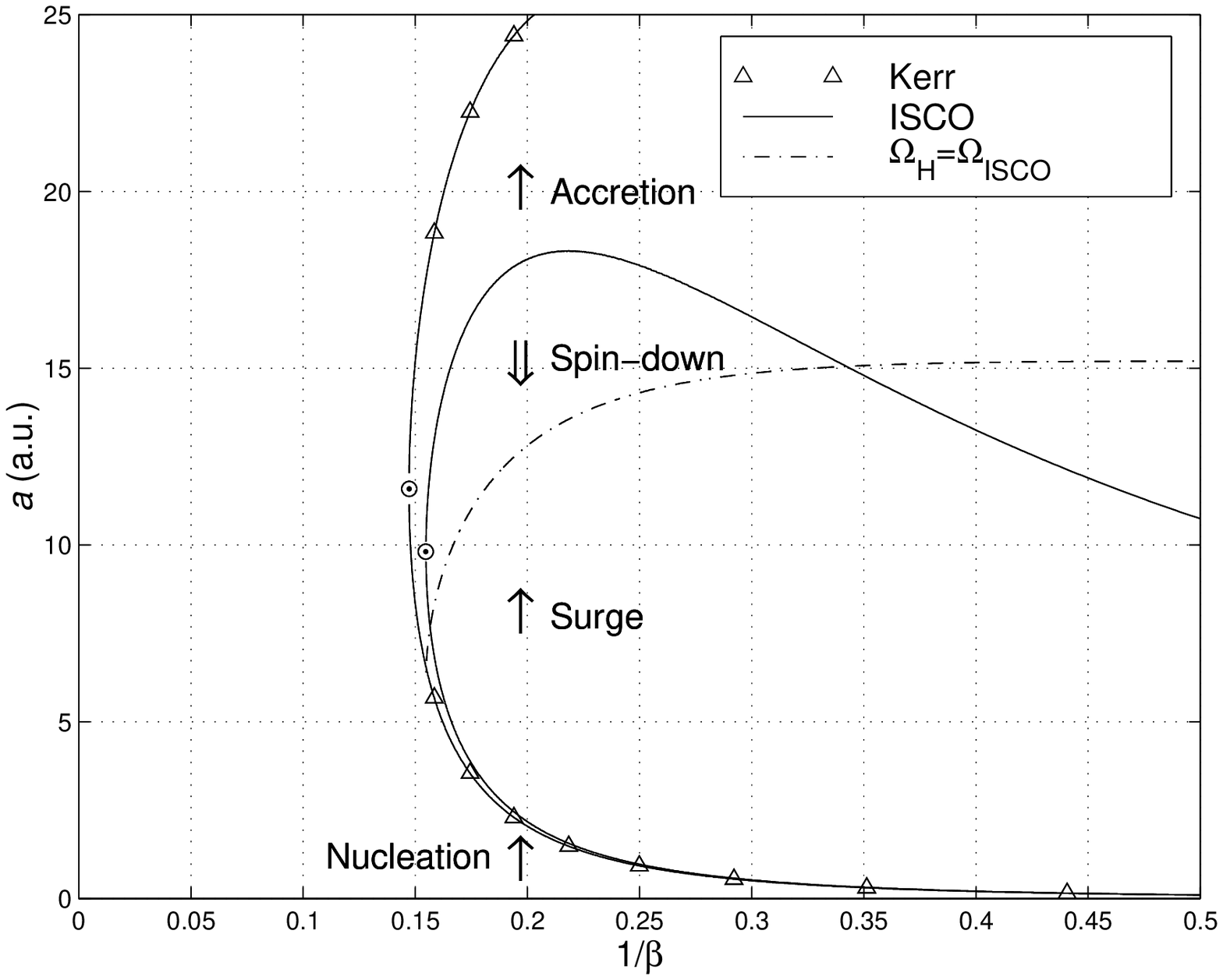}{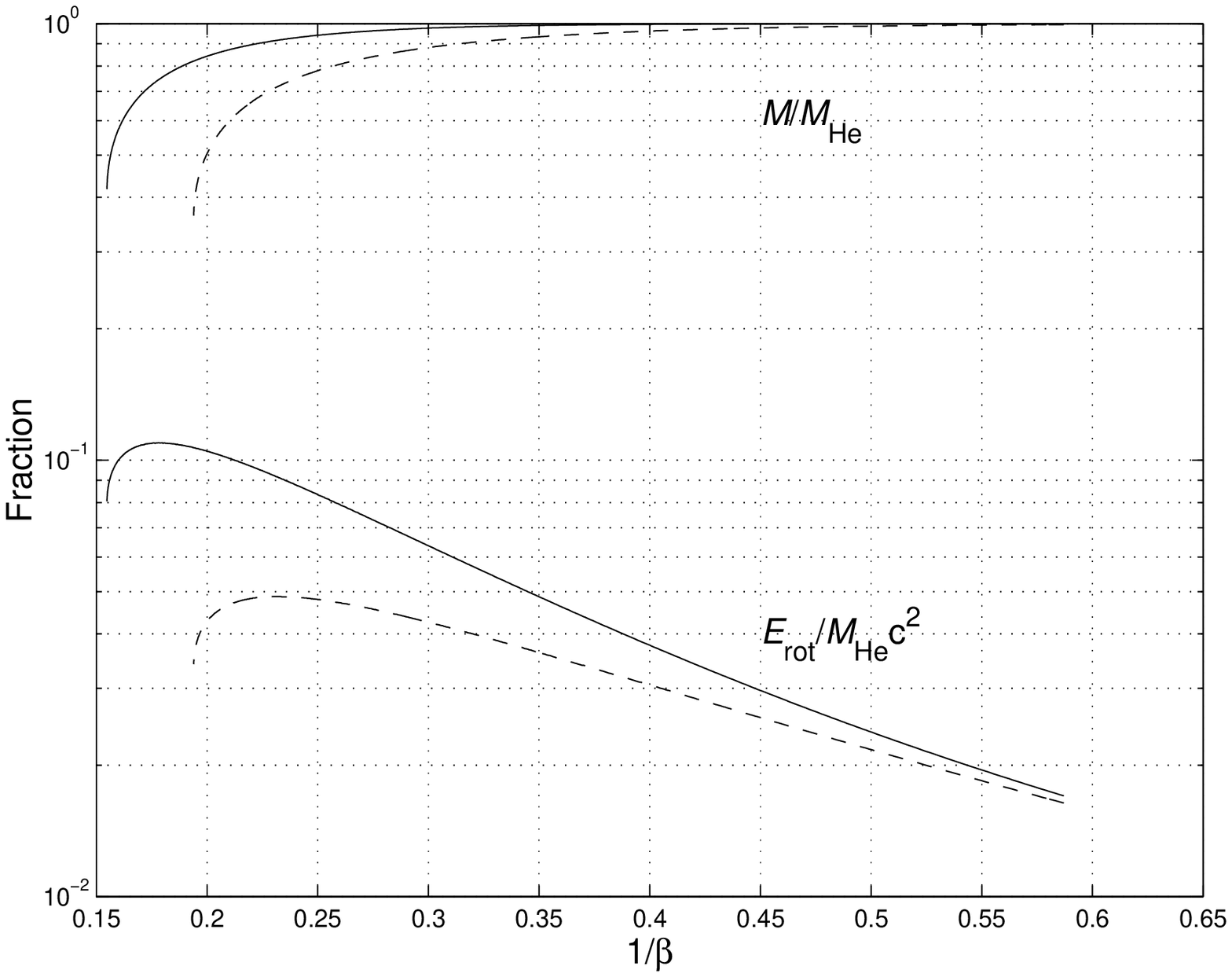}
\caption{}
\label{FIG_Y3}
\end{figure}

\begin{figure}
\plotone{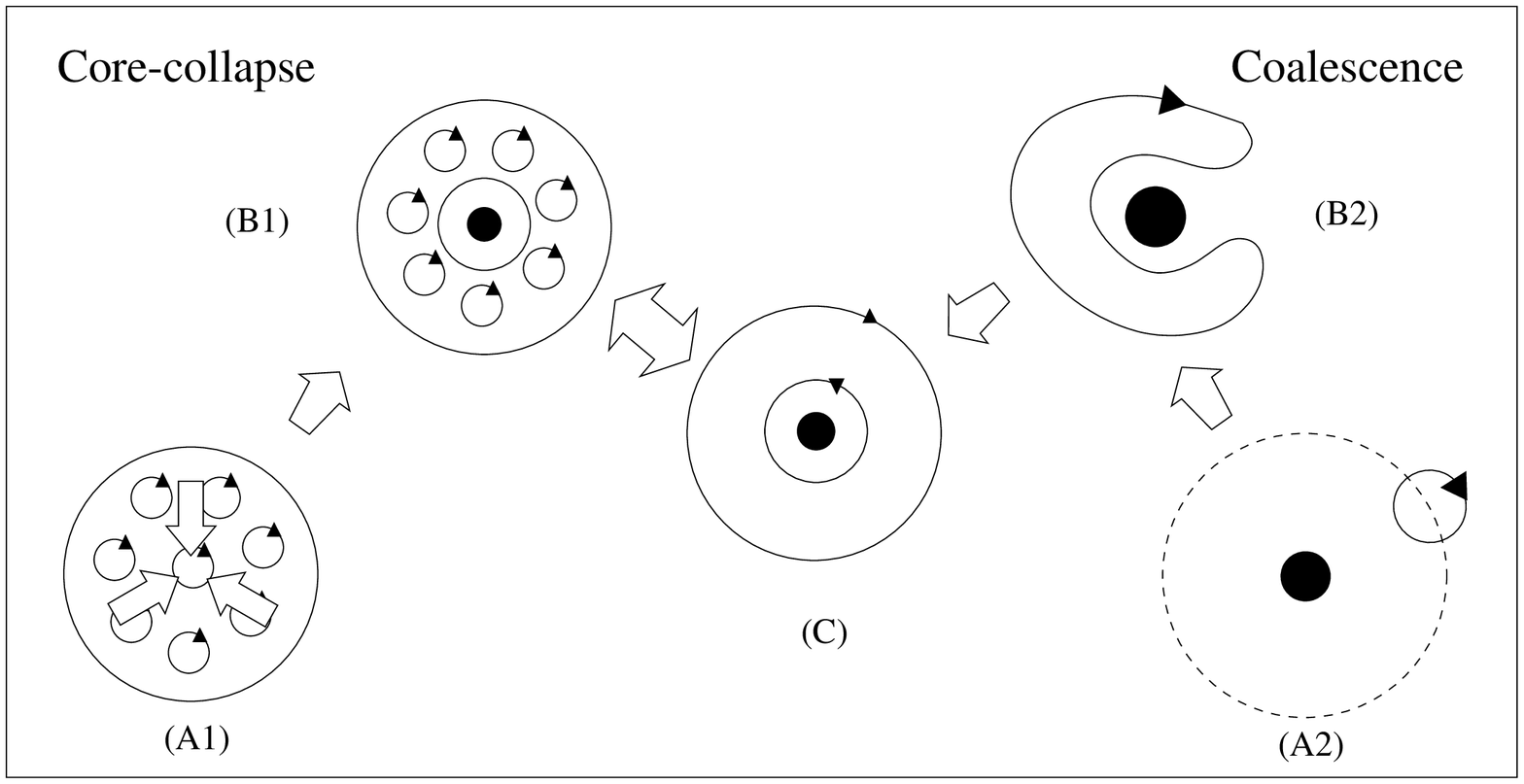}
\caption{}
\label{FIG_Y4}
\end{figure}

\begin{figure}
\plotone{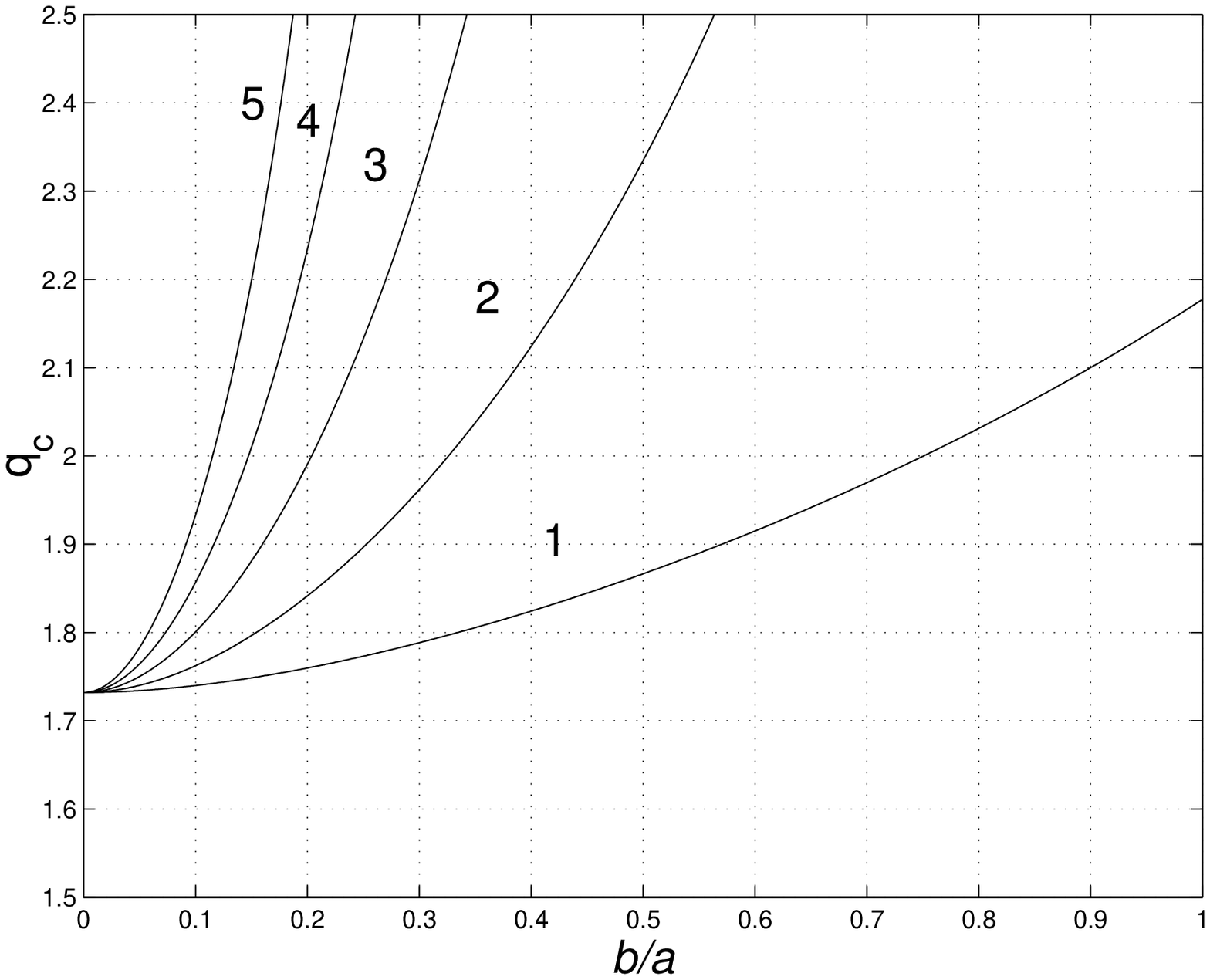}
\caption{}
\label{FIG_Y6}
\end{figure}

\begin{figure}
\plottwo{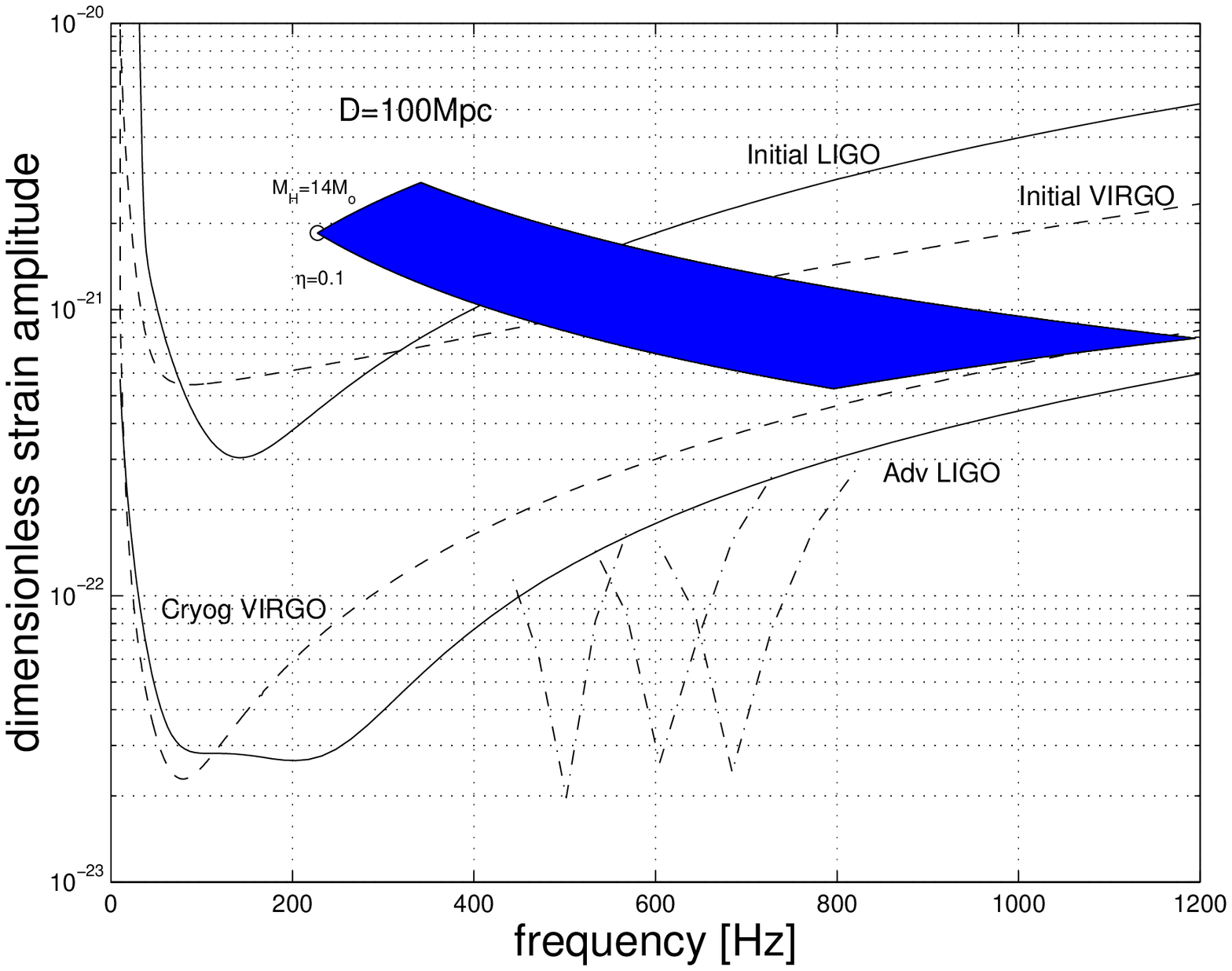}{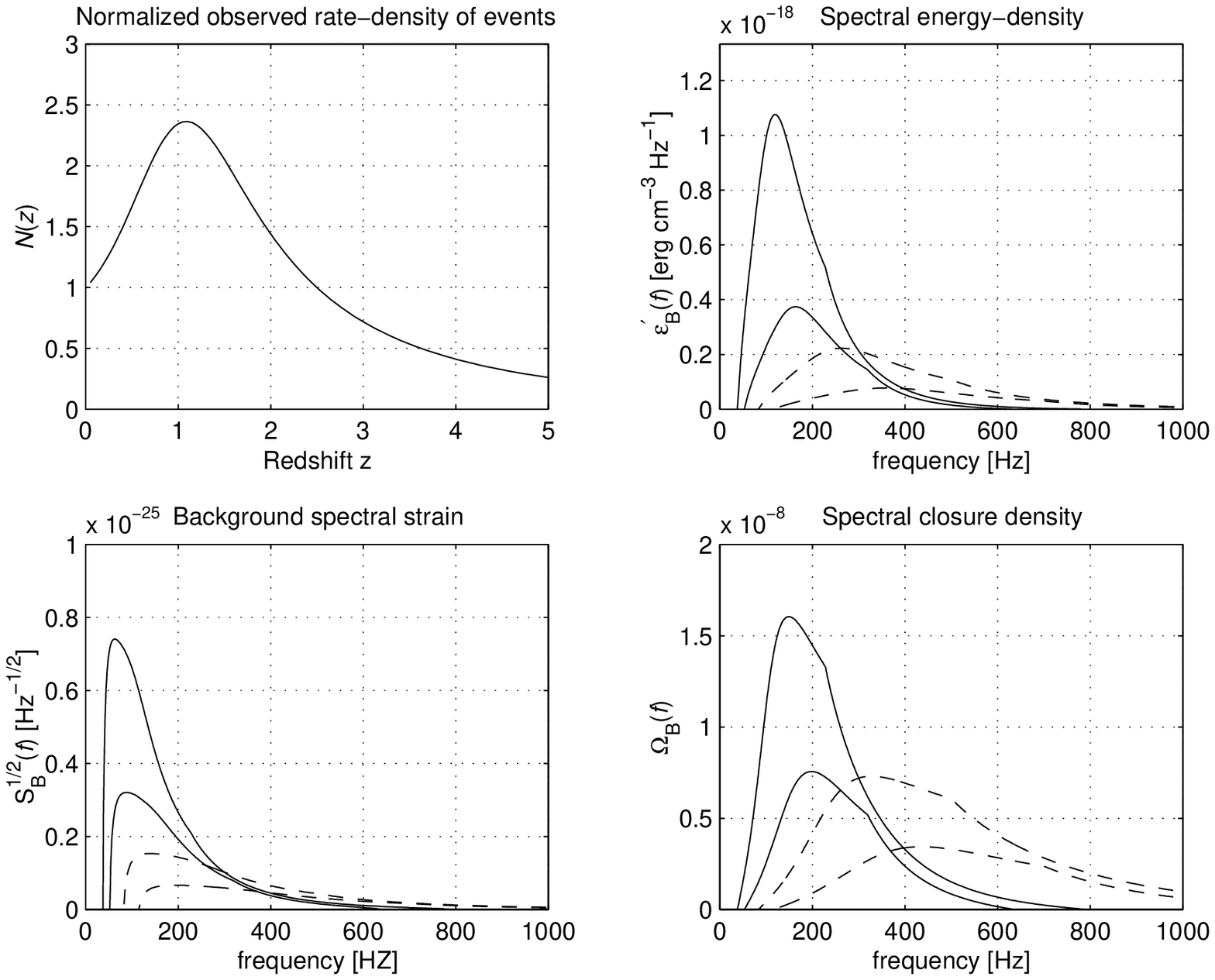}
\caption{}
\label{FIG_AREAL}
\end{figure}

\end{document}